\documentclass[apj]{emulateapj}
\usepackage{psfig}
\usepackage{color}

\def\simlt{\lower.5ex\hbox{$\; \buildrel < \over \sim \;$}}
\def\simgt{\lower.5ex\hbox{$\; \buildrel > \over \sim \;$}}

\def\hnot{\ifmmode H_0 \else H$_0$ \fi}

\def\msun{\ifmmode {\rm M_\odot} \else M$_\odot$\fi}
\def\lsun{\ifmmode {\rm L_\odot} \else L$_\odot$\fi}
\def\kms{km s$^{-1}$}
\def\deg{\ifmmode ^{\circ}
         \else $^{\circ}$\fi}
\def\pdeg{\ifmmode
           $\setbox0=\hbox{$^{\circ}$}\rlap{\hskip.11\wd0 .}$^{\circ}
     \else \setbox0=\hbox{$^{\circ}$}\rlap{\hskip.11\wd0 .}$^{\circ}$\fi}
\def\msunyr{\ifmmode {\rm M_\odot~yr^{-1}}\else${\rm M_\odot~yr^{-1}}$\fi}
\def\lam{\ifmmode {\lambda} \else {$\lambda$} \fi}
\def\lamLlam{\ifmmode \lambda L_{\lambda}(5100) \else {$\lambda L_{\lambda}(5100)$} \fi}
\def\nuLnu{\ifmmode \nu L_{\nu}(5100) \else {$\nu L_{\nu}(5100)$} \fi}

\def\mdoto{\ifmmode {\dot{M}_0} \else  $\dot{M}_0$ \fi}
\def\teff{\ifmmode {T_{eff}} \else $T_{eff}$ \fi}
\def\ilam{\ifmmode {I_\lambda} \else  $I_\lambda$ \fi}
\def\inu{\ifmmode {I_\nu} \else  $I_\nu$ \fi}
\def\fnu{\ifmmode {F_\nu} \else  $F_\nu$ \fi}
\def\yr{\ifmmode {\rm yr} \else  yr \fi}
\def\cm{\ifmmode {\rm cm} \else  cm \fi}
\def\cmmitwo{\ifmmode \rm cm^{-2} \else $\rm cm^{-2}$\fi}
\def\cmmithree{\ifmmode \rm cm^{-3} \else $\rm cm^{-3}$\fi}
\def\cmps{\ifmmode \rm cm~s^{-1}\else $\rm cm~s^{-1}$\fi}
\def\cmpsps{\ifmmode \rm cm~s^{-2}\else $\rm cm~s^{-2}$\fi}
\def\kmps{\ifmmode \rm km~s^{-1}\else $\rm km~s^{-1}$\fi}
\def\kmpspmpc{\ifmmode \rm km~s^{-1}~Mpc^{-1} \else
    $\rm km~s^{-1}~Mpc^{-1}$\fi}
\def\ergps{\ifmmode \rm erg~s^{-1} \else $\rm erg~s^{-1}$\fi}
\def\ergpspcm{\ifmmode \rm erg~s^{-1}~cm^{-2} \else $\rm erg~s^{-1}~cm^{-2}$ \fi}
\def\ergpspcmphz{\ifmmode \rm erg~s^{-1}~cm^{-2}~Hz^{-1} \else $\rm
erg~s^{-1}~cm^{-2}~Hz^{-1}$ \fi}
\def\ergpspcmpa{\ifmmode \rm erg~s^{-1}~cm^{-2}~\AA^{-1} \else $\rm
erg~s^{-1}~cm^{-2}~\AA^{-1}$ \fi}
\def\ergpsphz{\ifmmode \rm erg s^{-1} Hz^{-1} \else
   $\rm erg s^{-1} Hz^{-1}$ \fi}
\def\mbh{\ifmmode M_{\mathrm{BH}} \else $M_{\mathrm{BH}}$\fi}
\def\msigma{\ifmmode M_{\sigma} \else $M_{\sigma}$\fi}
\def\mbulge{\ifmmode M_{\mathrm{bulge}} \else $M_{\mathrm{bulge}}$\fi}
\def\lbulge{\ifmmode L_{\mathrm{bulge}} \else $L_{\mathrm{bulge}}$\fi}
\def\mgal{\ifmmode M_{\mathrm{gal}} \else $M_{\mathrm{gal}}$\fi}
\def\mgalstar{\ifmmode M^*_{\mathrm{gal}} \else $M^*_{\mathrm{gal}}$\fi}

\def\mbhsigstar{\ifmmode M_{\mathrm{BH}} - \sigma_* \else $M_{\mathrm{BH}} - \sigma_*$ \fi}
\def\mbhmbulge{\ifmmode M_{\mathrm{BH}} - M_{\mathrm{bulge}} \else $M_{\mathrm{BH}} - M_{\mathrm{bulge}}$ \fi}
\def\dmbh{\ifmmode \Delta~{\mathrm{log}}~M_{\mathrm{BH}} \else $\Delta$~log~$M_{\mathrm{BH}}$\fi}

\def\sigstar{\ifmmode \sigma_* \else $\sigma_*$\fi}
\def\signl{\ifmmode \sigma_{\mathrm{NL}} \else $\sigma_{\mathrm{NL}}$\fi}
\def\sigthree{\ifmmode \sigma_{\mathrm{[O~III]}} \else $\sigma_{\mathrm{[O~III]}}$\fi}
\def\sigtwo{\ifmmode \sigma_{\mathrm{[O~II]}} \else $\sigma_{\mathrm{[O~II]}}$\fi}
\def\wthree{\ifmmode {\rm FWHM({[O~III]})} \else $FWHM({[O~III]})$ \fi}
\def\wtwo{\ifmmode {\rm FWHM({[O~II]})} \else $FWHM({[O~II]})$ \fi}
\def\mthree{\ifmmode M_{\mathrm [O~III]} \else $M_{\mathrm [O~III]}$ \fi}
\def\mtwo{\ifmmode M_{\mathrm [O II]} \else $M_{\mathrm [O II]}$ \fi}
\def\hbeta{\ifmmode {\rm H}\beta \else H$\beta$\fi}
\def\mgii{\ifmmode {\rm Mg{\sc ii}} \else Mg~{\sc ii}\fi}

\def\lbreak{\ifmmode L_{\mathrm{break}} \else $L_{\mathrm{break}}$\fi}
\def\lcut{\ifmmode L_{\mathrm{cut}} \else $L_{\mathrm{cut}}$\fi}
\def\led{\ifmmode L_{\mathrm{Edd}} \else $L_{\mathrm{Edd}}$\fi}
\def\lbol{\ifmmode L_{\mathrm{bol}} \else $L_{\mathrm{bol}}$\fi}

\newcommand{\oiii}{{\sc [O~iii]}}
\newcommand{\oii}{{\sc [O~ii]}}
\newcommand{\sii}{{\sc [S~ii]}}
\newcommand{\feii}{Fe~{\sc ii}}

\slugcomment{Submitted to The Astrophysical Journal}

\shorttitle{The Black Hole - Galaxy Relationship for Quasars}
\shortauthors{Salviander \& Shields}

\begin{document}

\title{The Black Hole Mass - Stellar Velocity Dispersion Relationship for Quasars in the Sloan Digital Sky Survey Data Release 7}

\author{S. Salviander and G. A. Shields}
\affil{Department of Astronomy, University of Texas, Austin, TX 78712}

\begin{abstract}

We assess evolution in the \mbhsigstar\ relationship for quasars in the Sloan Digital Sky Survey Data Release 7 for the redshift range $0.1 < z < 1.2$. We estimate the black hole mass, \mbh, using the ``photoionization method,'' with the broad \hbeta\ or \mgii\ emission line and the quasar continuum luminosity. For the stellar velocity dispersion, we use the narrow \oiii\ or \oii\ emission line as a surrogate. This study is a follow-up to an earlier study in which we investigated evolution in the \mbhsigstar\ relationship in quasars from Data Release 3. The greatly increased number of quasars in our new sample has allowed us to break our lower-redshift subsample into black hole mass bins and probe the \mbhsigstar\ relationship for constant black hole mass. The \mbhsigstar\ relationship for the highest-mass ($\mbh > 10^{9.0}$ \msun) and lowest-mass ($\mbh < 10^{7.5}$ \msun) black holes appears to evolve significantly, however most or all of this apparent evolution can be accounted for by various observational biases due to intrinsic scatter in the relationship and to uncertainties in observed quantities. The \mbhsigstar\ relationship for black holes in the middle mass range ($10^{7.5} < \mbh < 10^{9.0}$ \msun) shows minimal change with redshift. The overall results suggest a limit of $\pm$0.2 dex on any evolution in the \mbhsigstar\ relationship for quasars out to $z \approx 1$ compared with the relationship observed in the local universe. Intrinsic scatter may also provide a plausible way to reconcile the wide range of results of several different studies of the black hole -- galaxy relationships. 

\end{abstract}


%% INTRO

\section{Introduction}\label{s:intro}

Strong correlations between the masses of central supermassive black holes and properties of host galaxies, coupled with the seeming ubiquity of such black holes, suggest an evolutionary link between black holes and their host galaxies. The two strongest correlations involve the mass of the black hole, \mbh, and either the host galaxy velocity dispersion, \sigstar\ (Gebhardt et al. 2000a; Ferrarese \& Merritt 2000), or the host galaxy luminosity (Magorrian et al. 1998) and therefore stellar mass. In an effort to determine whether the relationships hold for all masses and at all times, many studies have investigated these relationships for different lookback times, covering a wide range of black hole and galaxy properties. Over the last decade, the findings of these studies have spanned the range of possible results. Many works cite positive evolution, in the sense that black holes initially appear to outgrow their host galaxies (e.g. Peng et al. 2006; Shields et al. 2006; Salviander et al. 2007; Woo et al. 2008; Merloni et al. 2010; Decarli et al. 2010; Targett et al. 2012). Others find that black holes are comparatively smaller at high redshifts compared with their counterparts in the local universe (e.g. Alexander et al. 2008), or that there is no significant evolution in the relationship (e.g. Shields et al. 2003; Cisternas et al. 2011; Decarli et al. 2012). Various scenarios have been suggested for the physical origin of the black hole -- galaxy relationships, some of which include strong evolution (see Portinari et al. 2012 and references therein). A common scenario involves processes in which feedback from an active black hole couples with star-forming gas in the host galaxy (Silk
\& Rees 1998; Fabian 1999; King 2003; Di Matteo et al. 2005; Murray et al. 2005). However, it has been suggested that black hole -- galaxy relationships have no physical origin at all and can arise from a random process of mergers of galaxies with uncorrelated black hole mass and galaxy properties (Jahnke \& Macci{\`o} 2011).  

Observational bias, whether arising from intrinsic scatter or from uncertainties in observed properties, present a significant difficulty when attempting to assess evolution in any black hole -- galaxy relationship. Several observational biases have been identified and modeled (e.g. Salviander et al. 2007; Lauer et al. 2007; Shen \& Kelly 2010; Schulze \& Wisotzki 2011), many of which can mimic positive or negative evolution in black hole -- galaxy relationships where no intrinsic evolution exists. Volonteri \& Stark (2011) note that scatter and observational bias can also work to mask an intrinsic negative evolution. Lauer et al. (2007) warn that a false signal of evolution can be produced when comparing samples that have been selected on differing properties. For instance, local quiescent galaxy samples, which form the basis for the black hole -- galaxy correlations, are selected on galaxy properties, while evolutionary studies are often based on samples of active galactic nuclei (AGN), which are primarily selected on properties of the black hole. Comparison of black hole and galaxy properties between the two samples may yield misleading results. One way to mitigate this sort of bias is to compare lower- and higher-redshift samples selected on the same properties. 

This paper is a follow-up to a previous paper (Salviander et al. 2007) in which we used quasars from the Sloan Digital Sky Survey Data Release 3 (SDSS DR3) to assess evolution in the \mbhsigstar\ relationship in the redshift range $0.1 < z < 1.2$. In this paper, we expand our sample by including quasars from the more recent SDSS DR7. The greater number of quasars allows us to attempt to mitigate some of the effects of biases, particularly the comparison bias identified by Lauer et al., by assessing the evolution in \mbhsigstar\ over constant black hole mass. The \mbhsigstar\ relationship is an imperfect assessor of black hole -- galaxy evolution, as it appears to be a projection of a more fundamental black hole -- galaxy relationship involving the effective radius of the galaxy bulge (Hopkins et al. 2007); however, it is an accessible assessor, particularly for AGN in which surrogates from the AGN spectrum can be used in place of \sigstar. 

G\"{u}ltekin et al. (2009) give the fit to the local \mbhsigstar\ relationship as 
\begin{equation}
\label{e:msigmag}
\mbh = (10^{8.12}~\msun)(\sigma_e/200)^{4.24},
\end{equation}
where $\sigma_e$ is the effective velocity dispersion. However, for this paper, we adopt the earlier fit given by Tremaine et al. (2002) to facilitate comparison with our results from Salviander et al. (2007) (hereafter ``S07''). Tremaine et al. give the fit to the local relationship as 
\begin{equation}
\label{e:msigmat}
\mbh = (10^{8.13}~\msun)(\sigma_e/200)^{4.02},
\end{equation}

All values of luminosity used in this study are calculated using the cosmological parameters $\hnot = 70~\kmpspmpc, \Omega_{\rm M} = 0.3$, and $\Omega_{\Lambda} = 0.7$.

%% METHOD

\section{Method}\label{s:method}
\subsection{Black Hole Masses}

Our method for calculating black hole mass is described in S07 and Shields et al. (2003). Briefly, we use the ``photoionization method,'' which assumes that the broad line region (BLR) gas orbiting the black hole is virialized, such that $\mbh = fv^2R/G$. The factor $f$ represents a correction to the velocity field for an assumed geometry for the BLR. The radius of the BLR is derived from the radius-luminosity relationship, $R \propto L^{\gamma}$, calibrated by echo-mapping studies (Wandel et al. 1999; Kaspi et al. 2000, 2005). We adopt the formalism of Shields et al. (2003), who chose $\gamma = 0.5$, a choice motivated by photoionization physics and supported by the empirically determined radius-luminosity relationship of Bentz et al. (2006). The black hole mass is 
\begin{equation}
\label{e:mbh}
\mbh = (10^{7.69}~\msun)v_{3000}^2 L_{44}^{0.5},
\end{equation}
where the BLR velocity $v_{3000}$ is derived from the FWHM of the broad \hbeta\ or \mgii\ emission line in units of 3000 \kms\ and $L_{44}$ is the 5100 \AA\ continuum luminosity in units of $10^{44}$ \ergps. For higher redshifts we use the 4000 \AA\ continuum luminosity calibrated to a 5100 \AA\ luminosity assuming a power-law function fitted by Vanden Berk et al. (2001) for SDSS quasar composite spectra, $F_{\nu} \propto \nu^{\alpha_{\nu}}$ with $\alpha_{\nu} = -0.44$. We assume $f = \sqrt{3}/2$ for a flattened BLR geometry, which is incorporated into equation \ref{e:mbh}.

\subsection{Stellar Velocity Dispersion}

The stellar velocity dispersion, \sigstar, is difficult to measure directly from the spectra of quasars, so we use emission line surrogates from the narrow line region (NLR) of the quasar. If the NLR gas orbits in the gravitational potential of the host galaxy bulge, this will produce a one-to-one proportionality between \sigstar\ and the width of narrow emission lines. Greene \& Ho (2005) examine correlations between \sigstar\ and the widths of \oiii, \oii, and \sii\ for a sample of narrow-line AGN, and find that \sigthree\ tends to be wider than \sigstar. This is likely due to the proximity of the \oiii-emitting gas to the active nucleus producing emission in the blue wing of \oiii. Nelson \& Whittle (1996) also note that blue wings tend to be present on \oiii\ emission lines. However, work done by Nelson (2000) and Bonning et al. (2005) show that in the mean the width of the $\lambda$5007 \oiii\ emission line tracks \sigstar\ for a range of AGN luminosities. Greene \& Ho find that the lower-ionization lines \oii\ and \sii\ track \sigstar\ in the mean, though with considerable scatter. The correlation between \oiii\ and \oii\ width shown in Figure 4 of S07 further suggests that \oii\ may be used as a surrogate for \sigstar\ with some confidence. We define the surrogate \sigstar\ as $\signl = $ FWHM(\oiii)$/2.35$ or FWHM(\oii)$/2.35$ for a Gaussian profile.   

%% SAMPLE

\section{Sample Selection and Spectrum Measurements}\label{s:sample}

The quasars for this study were taken from the SDSS DR7 (Abazajian et al. 2009). Sample selection and measurements were carried out as described in S07, except as noted below. Briefly, we selected all objects from DR7 that were classified by the Spectroscopic Query Form as quasars. We restricted our search to the redshift range $0.1 \leq z \leq 1.4$; the lower limit minimizes the inclusion of Seyfert galaxies, while the upper limit is the maximum redshift at which the \oii\ emission line cleanly appears in the SDSS spectral window.

We carried out spectrum corrections and measurements using an automated fitting algorithm that first corrects the spectra for galactic extinction and rebins the wavelength scale from logarithmic to linear with 1.41 \AA\ pixel$^{-1}$. We subtracted off the optical and UV \feii\ using templates from Marziani et al. (2003) for the optical regime and a reconstructed template from Vestergaard \& Wilkes (2001) augmented by theoretical data from Sigut \& Pradhan (2003) for the UV regime. The line flux, velocity dispersion ($\sigma_{\mathrm{GH}}$), and FWHM of \hbeta, \oiii, \oii, and \mgii\ are measured both before and after subtraction of \feii. The emission lines are modeled using Gauss-Hermite functions in order to account for any asymmetries (characterized by the $h_3$ parameter) and deviations from Gaussianity (characterized by the $h_4$ parameter). A special case is the FWHM of the \oii\ $\lambda\lambda$3726,3729 doublet, which tends to be unresolved in SDSS spectra. We follow the procedure of S07, in which we found the \oii\ doublet to be successfully modeled as a single line from which the intrinsic width is inferred using a calibration curve. Continuum fluxes are measured at 4000 \AA\ and 5100 \AA\ in the observed frame. The instrumental width is subtracted in quadrature from the measured line widths, and all measurements are corrected to the rest frame.

We created two subsamples as described in S07: a lower-redshift sample using \hbeta\ and \oiii\ in the redshift range $0.1 \leq z \leq 0.81$ (hereafter the ``HO3'' sample) and a higher-redshift sample using \mgii\ and \oii\ in the redshift range $0.4 \leq z \leq 1.35$ (hereafter the ``MO2'' sample). The HO3 and MO2 samples consist of 5355 and 808 objects, respectively. 

We executed a series of quality cuts to remove objects with substandard spectra as described in S07 with the exception of relaxed standards for the EW and FWHM errors when it was found through visual inspection that our prior numerical cuts eliminated a number of usable spectra. Briefly, we eliminated objects on the basis of EW and FWHM measurement errors, excessive deviation from Gaussianity (the $h_4$ parameter), reduced $\chi^2$, and a minimum width for the broad lines. Table \ref{t:ecuts} shows the quality criteria implemented for each emission line. We determined the new EW and FWHM error cuts by selecting a random subsample of 100 DR7 spectra, eliminating poor quality spectra through visual inspection alone, and adjusting the numerical cuts until we achieved a very good match with the hand-selected subsample. There are 25,009 DR7 quasars with $0.1 \leq z \leq 0.81$ and 38,811 quasars in the range $0.4 \leq z \leq 1.4$. Table \ref{t:cuts} shows the number of objects remaining in the subsamples after each quality cut was implemented. The greatest reduction in the number of objects for the MO2 sample results from the absence of the \oii\ emission line from the spectra. A great reduction for both subsamples occurs with the EW \% error cut, which tends to eliminate objects with the poorest signal-to-noise ratios. Most of the objects eliminated by the EW \% error cut were based on poor S/N for \hbeta\ or \oii. 

We found in S07 that radio loudness did not significantly affect our results, so we have made no effort to exclude radio-loud objects from this sample. 

\begin{deluxetable}{lcccc}
\tablewidth{0pt}
\tablecaption{Quality Cut Tolerances for Each Emission Line\label{t:ecuts}}
\tablehead{
\colhead{Quality cut} & 
\colhead{\hbeta} &
\colhead{\oiii} &
\colhead{\mgii} &
\colhead{\oii}} 
\startdata
EW error & 7\% & 5\% & 5\% & 11\% \\
FWHM error & 15\% & 15\% & 15\% & 30\% \\
profile shape ($h_4$) & 0.2 & 0.2 & 0.2 & 0.2 \\
Reduced $\chi^2$ & 4 & 4 & 4 & 4 \\
BL minimum width (\kmps) & 1500 & -- & 1500 & -- \\
\enddata
%\tablecomments{See text for discussion.}
\end{deluxetable}

\begin{deluxetable}{lcc}
\tablewidth{0pt}
\tablecaption{Quality Cut Reductions for the Subsamples\label{t:cuts}}
\tablehead{
\colhead{Quality cut} &
\multicolumn{2}{c}{\# of objects remaining} \\
\colhead{} &
\colhead{HO3} &
\colhead{MO2}}
\startdata
Initial sample & 22,985 & 13,213 \\
EW error & 14,674 & 2214 \\
FWHM error & 12,270 & 2065 \\
Profile shape ($h_4$) & 11,116 & 1334 \\
Reduced $\chi^2$ & 10,899 & 1322 \\
BL width minimum & 10,240 & 1303 \\
Visual inspection & 5355 & 808 \\
\enddata
%\tablecomments{See text for discussion.}
\end{deluxetable}

%% RESULTS

\section{Results}\label{s:results}
\subsection{The \mbhsigstar Relationship}\label{s:mbhsigstar}

\begin{figure}[htbp]
\begin{center}
\includegraphics[scale=0.4]{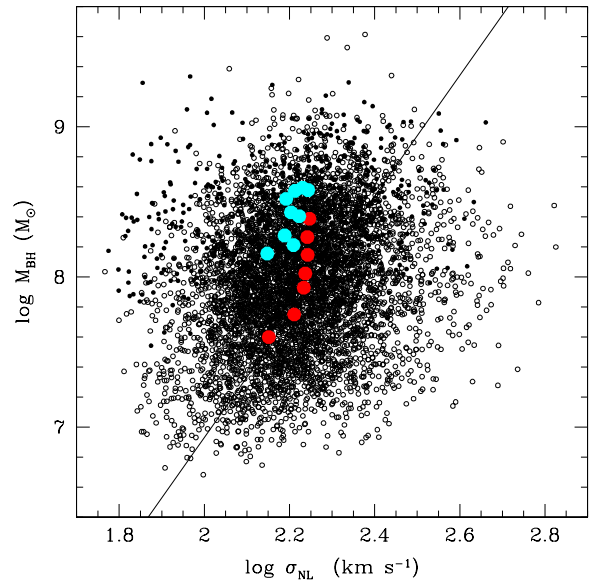}
\caption{The \mbhsigstar\ relation for the HO3 (small open circles) and MO2 (small closed circles) samples. The large circles show the average \mbh\ and \sigthree\ for the redshift bins shown in Table \ref{t:zbins}---red for HO3 and cyan for MO2. The line is not a fit to data, but rather the local \mbhsigstar\ relationship as given by Tremaine et al. (2002).}\label{f:msigma}
\end{center}
\end{figure}

Figure \ref{f:msigma} shows the \mbhsigstar\ relationship for our sample of quasars for both the HO3 and MO2 samples. The large red and cyan circles show the mean \mbh\ and \signl\ (\sigthree\ or \sigtwo, respectively) for redshift bins, discussed in the following section. The solid line represents the local \mbhsigstar\ relationship given by equation \ref{e:msigmat} and is not a fit to the data. The dispersion with respect to the local relationship shown in Figure \ref{f:msigma} is 0.63 dex and 0.72 dex for the HO3 and MO2 samples, respectively. These are consistent with the dispersions for the subsamples in S07.

The scatter is mostly attributable to the dispersion in the \signl-\sigstar\ surrogacy and the uncertainty in the virial estimate for \mbh\ (see \S \ref{s:biases}). Even with the high degree of scatter evident in Figure \ref{f:msigma}, the mean \mbh\ and \signl\ do center on the local \mbhsigstar\ relationship for $z < 0.45$. For greater redshifts, the narrow-line width saturates at log $\signl \sim 2.24$, and the means become increasingly displaced above the relationship. 

% msigma

\subsection{Evolution with Lookback Time}\label{s:zdep}

We compare the photoionization mass with the ``narrow-line mass,'' that is, the black hole mass calculated with equation \ref{e:mbh} using \sigthree\ or \sigtwo\ in place of \sigstar\ ($M_{\sigma}$). The displacement from the local \mbhsigstar\ relationship is defined as $\Delta {\rm log \ } \mbh \equiv {\rm log \ } \mbh - {\rm log \ } M_{\sigma}$. If a quasar follows the local \mbhsigstar\ relationship, it will show perfect agreement between the photoionization mass and narrow-line mass, i.e. $\dmbh = 0$. If $\dmbh > 0$, this indicates the black hole is proportionally more massive than would be expected from \signl\ and the local \mbhsigstar\ relationship; likewise, if $\dmbh < 0$, this indicates the black hole is proportionally less massive than expected from \signl\ and the local relationship. The mean \dmbh\ for the HO3 sample is $+0.13$ dex, which heavily weights the abundant low-$z$ quasars in the sample; the mean \dmbh\ for MO2 is $+0.59$. Both indicate black holes that are proportionally more massive than expected. These are consistent with the subsample means of S07. 

% dmbh
\begin{figure}[htbp]
\begin{center}
\includegraphics[scale=0.4]{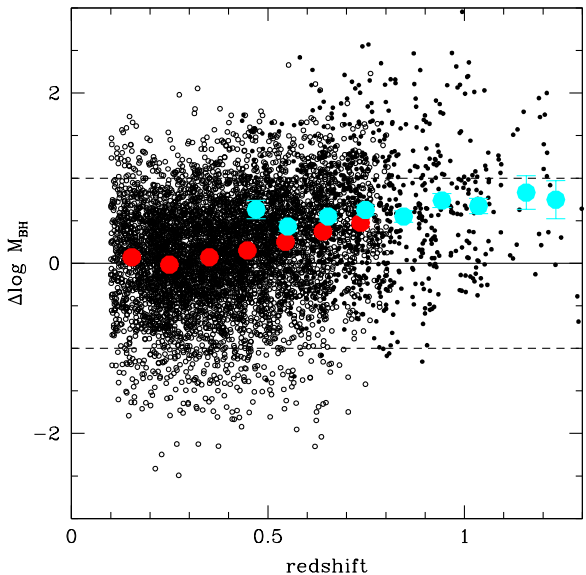}
\caption{\dmbh---offset from the local \mbhsigstar\ relation---as a function of redshift for the HO3 (small open circles) and MO2 (small closed circles) samples. The large circles show the average \dmbh\ for the redshift bins shown in Table \ref{t:zbins}---red for HO3 and cyan for MO2. For most of the redshift bins, the error bars are smaller than the marker.}\label{f:dmbh}
\end{center}
\end{figure}

\begin{deluxetable*}{lcccccc}
\tablewidth{0pt}
\tablecaption{Average Quantities for Redshift Bins\label{t:zbins}}
\tablehead{
\colhead{$z$ (Number of Objects)} & 
\colhead{\nuLnu} & 
\colhead{\signl} & 
\colhead{\mbh} & 
\colhead{\dmbh} & 
\colhead{\led/\lbol} & 
\colhead{FWHM$_{\mathrm{BL}}$} \\
\colhead{} &
\colhead{\ergps} &
\colhead{\kmps} &
\colhead{\msun} &
\colhead{} & 
\colhead{} & 
\colhead{\kmps}}
\startdata
\multicolumn{7}{c}{HO3 Sample} \\
\hline 
0.15  (685) & 43.74 & 2.15 & 7.60 & $+$0.07 & $-$1.01 & 3.54 \\
0.25 (1153) & 43.98 & 2.21 & 7.75 & $-$0.02 & $-$0.92 & 3.55 \\
0.35 (1171) & 44.24 & 2.23 & 7.92 & $+$0.07 & $-$0.84 & 3.58 \\
0.45 (1030) & 44.41 & 2.24 & 8.02 & $+$0.15 & $-$0.76 & 3.58 \\
0.54  (686) & 44.60 & 2.24 & 8.15 & $+$0.25 & $-$0.69 & 3.58 \\
0.64  (419) & 44.76 & 2.24 & 8.26 & $+$0.37 & $-$0.65 & 3.60 \\
0.74  (201) & 44.98 & 2.25 & 8.39 & $+$0.48 & $-$0.55 & 3.60 \\
\hline 
\multicolumn{7}{c}{MO2 Sample} \\
\hline 
0.47  (43) & 44.62 & 2.15 & 8.16 & $+$0.63 & $-$0.69 & 3.59 \\
0.55 (117) & 44.78 & 2.22 & 8.22 & $+$0.43 & $-$0.59 & 3.57 \\
0.65 (166) & 44.92 & 2.20 & 8.27 & $+$0.55 & $-$0.50 & 3.56 \\
0.75 (177) & 45.02 & 2.22 & 8.41 & $+$0.62 & $-$0.54 & 3.61 \\
0.85 (129) & 45.10 & 2.23 & 8.40 & $+$0.55 & $-$0.45 & 3.59 \\
0.94  (87) & 45.26 & 2.22 & 8.56 & $+$0.74 & $-$0.44 & 3.62 \\
1.04  (61) & 45.35 & 2.25 & 8.59 & $+$0.67 & $-$0.38 & 3.61 \\
1.16  (14) & 45.29 & 2.19 & 8.49 & $+$0.83 & $-$0.35 & 3.59 \\
1.23  (13) & 45.55 & 2.23 & 8.59 & $+$0.75 & $-$0.19 & 3.57 \\
\enddata
\tablecomments{All quantities are expressed in logarithmic units except for redshift. \signl\ corresponds to \sigthree\ for the HO3 sample and to \sigtwo\ for the MO2 sample; FWHM$_{\mathrm{BL}}$ corresponds to the \hbeta\ FWHM for the HO3 sample and to the \mgii\ FWHM for the MO2 sample. Bins with fewer than ten objects were excluded (this affected the MO2 sample only).}
\end{deluxetable*}

The pivotal question is whether the \mbhsigstar\ relationship evolves with lookback time. In Figure \ref{f:dmbh} we plot \dmbh\ as a function of redshift for the HO3 and MO2 samples, including the mean for redshift bins incremented by $\Delta z = 0.1$. Table \ref{t:zbins} shows various quantities for the redshift bins for HO3 and MO2. There is a general upward trend in \dmbh\ with redshift, with an overall increase in the average \dmbh\ of $+0.39$ dex from $z = 0.1$ to $z = 0.8$ for HO3 only and $+0.67$ dex from $z = 0.1$ to $z \approx 1.2$ for both HO3 and MO2. Figure \ref{f:dmbh} is consistent with our result from S07.

% msigma bins
\begin{figure}[htbp]
\begin{center}
\includegraphics[scale=0.4]{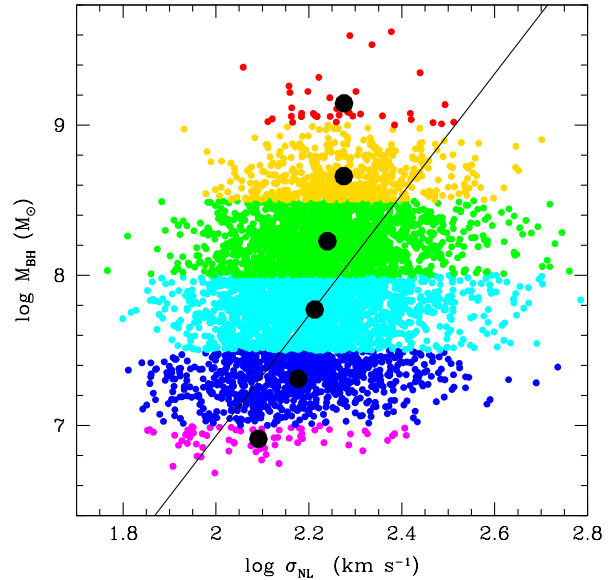}
\caption{The \mbhsigstar\ relation for the HO3 sample, with colors corresponding to the mass bins shown in Figure \ref{f:dmbhbins}. The large black circles show the average \mbh\ and \sigthree\ for the mass bins. The line is not a fit to data, but rather the local \mbhsigstar\ relationship as given by Tremaine et al. (2002).}\label{f:msigmabins}
\end{center}
\end{figure}

\begin{figure}[htbp]
\begin{center}
\includegraphics[scale=0.4]{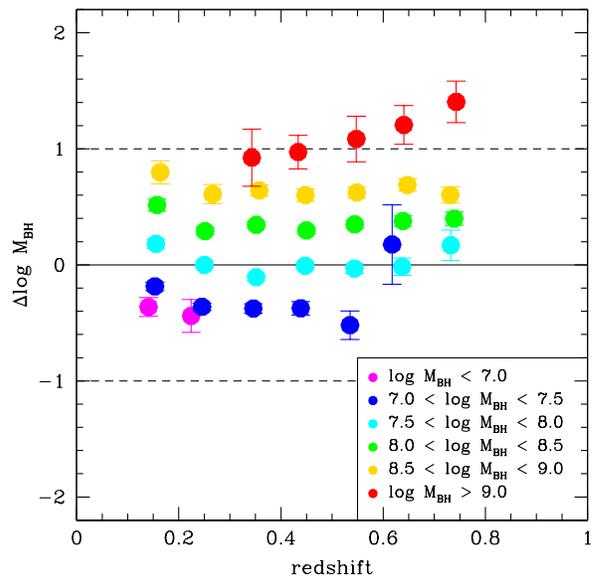}
\caption{\dmbh---offset from the local \mbhsigstar\ relation---as a function of redshift for the black hole mass bins. % comment on blue outlier point)
}\label{f:dmbhbins}
\end{center}
\end{figure}

Does this upward trend represent real evolution in the \mbhsigstar\ relationship with lookback time or is it partially or wholly the result of other effects contributing to the appearance of evolution? We discuss potential biases in \S \ref{s:biases}. An immediate test is to separate our sample into bins restricted by \mbh\ and determine whether there is evidence for evolution with lookback time within each of these mass bins. The limited number of objects in our sample for S07 prohibited such a test for the higher mass bins, but with the inclusion of quasars from DR7 we have sufficient objects to separate the current HO3 sample into six mass bins, ranging from $\mathrm{log\ } \mbh < 7.0$ \msun\ to $\mathrm{log\ } \mbh > 9.0$ \msun, that are incremented by 0.5 dex \msun. (There are too few objects in our MO2 sample to break it into comparable mass bins.) Figure \ref{f:msigmabins} shows the \mbhsigstar\ relation color-coded according to the mass bins. Figure \ref{f:dmbhbins} shows the same comparison as Figure \ref{f:dmbh} using average quantities for each of the mass bins, and shows how the intrinsic scatter in the \mbhsigstar\ relationship leads to systematic offsets from the local relationship that scale with black hole mass. Note the apparent lack of significant evolution in the \mbhsigstar\ relationship except for the highest and lowest black hole masses.

% dmbh bins

It is likely that the overall appearance of an upward trend in \dmbh\ with redshift in Figure \ref{f:dmbh} is the result of a combination of different selection biases. We discuss these in the following section. 

%% BIASES

\section{Uncertainties and Biases}\label{s:biases}

We now discuss potential selection biases that could lead to the appearance of evolution in the \mbhsigstar\ relationship with redshift. At the heart of these selection biases are scatter and measurement uncertainty. G\"{u}ltekin et al. (2009) determined an intrinsic rms scatter in the \mbhsigstar\ relation---that is, scatter not due to uncertainties in measurements and surrogacies---of 0.31 dex for early-type galaxies and 0.44 dex for all galaxy types, based on the locally-observed sample of galaxies with dynamically-determined black hole masses. For AGN samples, neither the magnitude of the intrinsic scatter in the \mbhsigstar\ relationship nor the shape of the scatter function is known. The 0.63 and 0.72 dex scatter about the \mbhsigstar\ relation for our HO3 and MO2 samples, respectively, includes the intrinsic scatter in the \mbhsigstar\ relation, the $\sim0.4$ dex uncertainty for virial estimates of black hole mass (Vestergaard \& Peterson 2006), and the 0.13 dex scatter in the \oiii-\sigstar\ surrogacy (Bonning et al. 2005). The combined scatter and uncertainty leads to the systematic offsets for the mass bins shown in Figure \ref{f:dmbhbins}, with the magnitude of the offsets proportional to the scatter. The scatter and uncertainties factor into several biases, as discussed below. 

\subsection{Malmquist-like bias}

In S07 we simulated the effect of a Malmquist-like bias that arises from correlations between quasar luminosity, \mbh, and redshift. Galaxies hosting larger black holes tend to have higher quasar luminosities and are thus preferentially selected from the flux-limited SDSS sample. Because there is intrinsic scatter in the \mbhsigstar\ relation, a given galaxy in a flux-limited quasar sample will typically host a disproportionately large black hole. Assuming that quasar luminosity is correlated with \mbh, the correlation between luminosity and redshift means this selection effect is increased with redshift, and therefore leads to a positive trend in \dmbh\ with redshift. The strength of this trend is related to the magnitude of the scatter. Since it is not known whether the intrinsic scatter (both the magnitude and the shape of the scatter function) in the \mbhsigstar\ relationship for quasars is the same as that for locally-observed galaxies, the degree of bias is also not known. Qualitatively, Figure \ref{f:dmbhbins} indicates that this bias is operating in the lowest mass bin, in which lower-luminosity objects drop out of the sample with increasing redshift (see also Figure \ref{f:lobin}). In our simulations of the Malmquist-like bias, scatters of 0.3 dex and 0.5 dex produced a factor of 0.19 and 0.44 dex bias, respectively, in \mbh\ for a sample with an average mass of log $\mbh \approx 8.0$, roughly corresponding to the average black hole mass of the HO3 sample. This range of bias is consistent with the overall rise in \dmbh\ of $+0.39$ dex for the HO3 sample (see Figure \ref{f:dmbh}). 

\subsection{Lauer bias}

\begin{figure}[htbp]
\begin{center}
\includegraphics[scale=0.4]{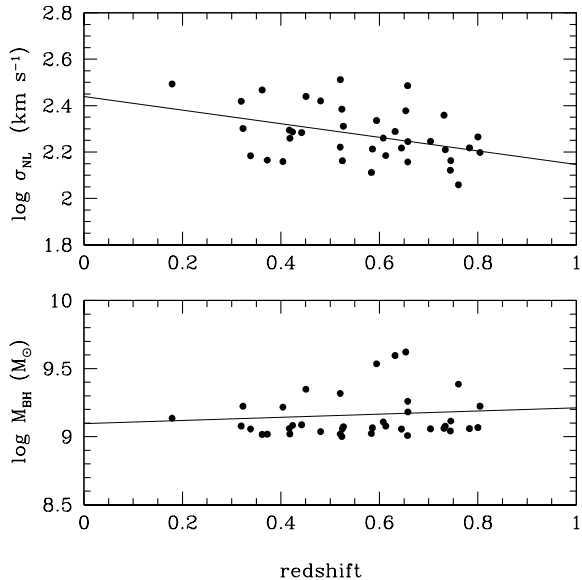}
\caption{Log \signl\ as a function of redshift (top panel) and log \mbh\ as function of redshift (bottom panel) for the highest-mass (log \mbh $> 9.0$ \msun) HO3 bin. See text for discussion.}\label{f:hibin}
\end{center}
\end{figure}

Lauer et al. (2007) discuss how the appearance of evolution can arise from comparison of high-redshift samples of AGN with local samples of quiescent galaxies, which have differing selection criteria. Bias in the \mbhsigstar\ relationship manifests at high \mbh\ where there is a dearth of high-\sigstar\ galaxies to host the biggest black holes. Their Figure 3 shows how intrinsic scatter in the \mbhsigstar\ relationship gives rise to a higher density of the most massive black holes than for the ``native'' population hosted by galaxies with \sigstar\ prescribed by the \mbhsigstar\ relationship. This leaves galaxies with more modest \sigstar\ as the only remaining homes for the majority of the most massive black holes. Such a bias may account for the rise in \dmbh\ for the highest-mass bin. Lauer et al. estimate, for reasonably large scatter, a factor of 3 bias in \mbh\ for the \mbhsigstar\ relation for the biggest black holes, consistent with the rise in the largest mass bin in Figure \ref{f:dmbhbins}. Figure \ref{f:hibin} shows that, while the black hole mass is roughly constant with redshift for this bin, \sigthree\ decreases with redshift. A possible explanation for this drop off in \sigthree\ with redshift is evolution in the \sigstar\ function. Chae (2011) compares \sigstar\ functions at $z=0$ and $z=1$, based on galaxy data drawn from SDSS DR5 and DR6, and shows not only a dramatic decrease in large-\sigstar\ galaxies at $z=1$ compared with $z=0$, but an increase in the number density of small-\sigstar\ galaxies. Combined with a \sigstar-function that already drops off more rapidly than the black hole mass function, this leaves increasingly modest-\sigstar\ galaxies to host most of the biggest black holes with redshift.

% hibin

\subsection{Shen-Kelly bias}

Shen \& Kelly (2010) explore a bias that arises from uncertainty in the calculation of \mbh\ using the photoionization method. Uncertainties in the photoionization mass arise from, among other quantities, uncertainty in the quasar luminosity. Scatter in the luminosity coupled with the greater abundance of low-mass black holes in the black hole mass function means that more of the modest black holes are scattered to higher luminosities than the other way around. This results in a greater number of black holes observed to have higher masses. These black holes with observed masses greater than their true masses are nonetheless observed to reside in galaxies with true \sigstar\ corresponding to the true \mbh, since the measurements (or proxies) for \sigstar\ are not subject to the same selection biases. Such a bias would give rise to the appearance of evolution in the \mbhsigstar\ relationship. Shen \& Kelly estimate a 0.2 - 0.3 dex bias in \mbh\ for $\lbol > 10^{46}$ \ergps, corresponding to the average bolometric luminosity for our highest-mass bin. This black hole mass bias accounts for at least half of the observed offset in this bin at $z \sim 0.8$. Since this bias operates independently of the Lauer bias, it would represent an additional contribution to the upward trend for the highest mass bin in Figure \ref{f:dmbhbins}. 

\subsection{Schulze-Wisotzki bias}

\begin{figure}[htbp]
\begin{center}
\includegraphics[scale=0.4]{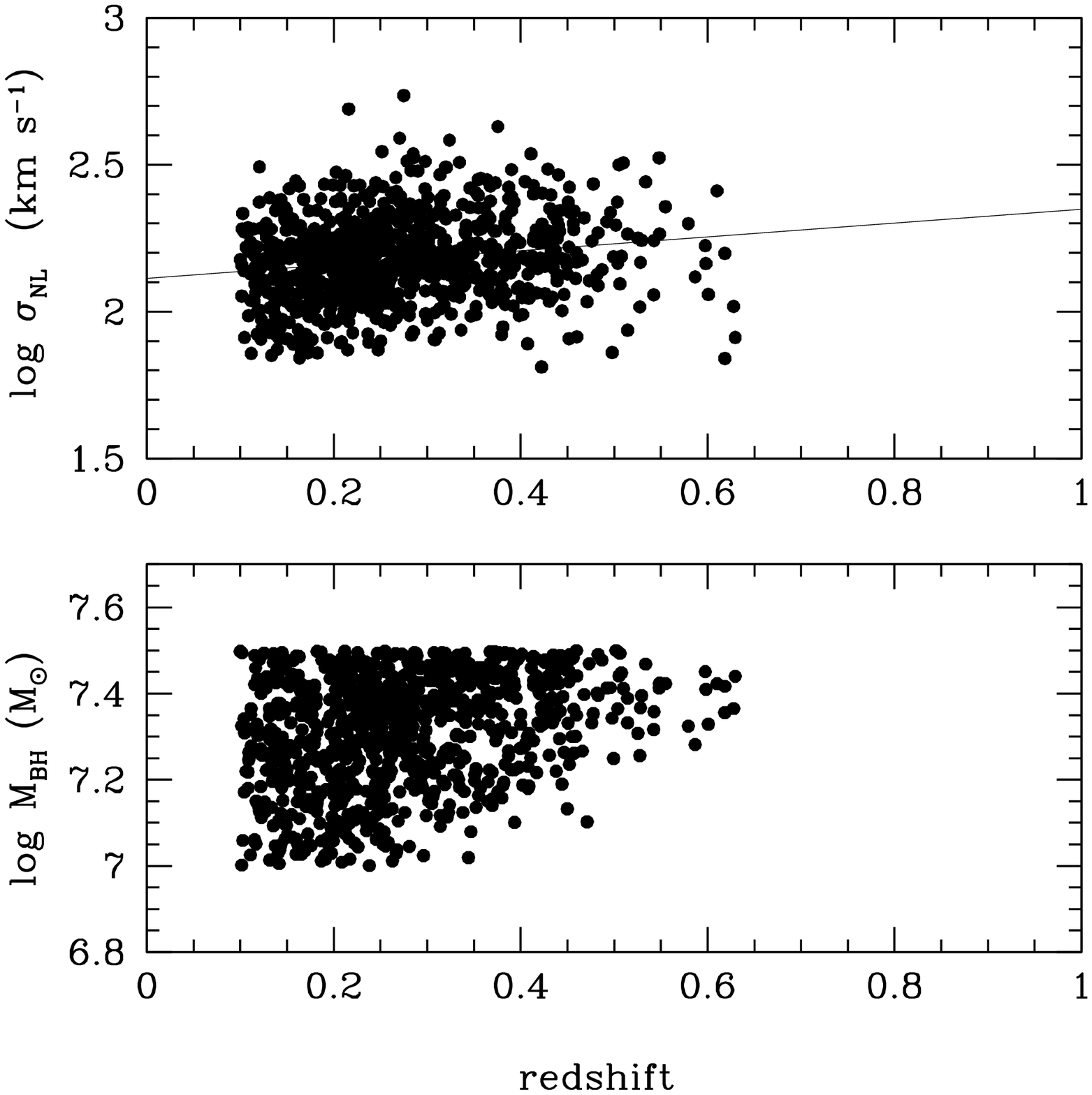}
\caption{Log \signl\ as a function of redshift (top panel) and log \mbh\ as function of redshift (bottom panel) for the $7.0 < \mathrm{log \ } \mbh < 7.5$ \msun\ HO3 bin. See text for discussion.}\label{f:lobin}
\end{center}
\end{figure}

Schulze \& Wisotzki (2011) investigate several biases that could affect studies of the evolution in the black hole - galaxy relationship in AGN, including the heretofore unexplored active fraction bias. Scatter in the black hole - galaxy relationship coupled with an active fraction that depends on black hole mass can produce either a positive or negative bias in a luminosity-limited sample, depending on whether the active fraction increases or decreases with \mbh. A negative bias occurs if the active fraction is inversely proportional to \mbh. In such a case, smaller black holes have a greater probability of being fueled, and therefore detected in a sample, than larger black holes, producing an AGN sample that is more likely to include comparatively smaller black holes for a given host galaxy property. Schulze \& Wisotzki found the active fraction bias in a sample like the SDSS to be strongest at the luminosity limit of the survey, thus affecting the smallest black holes. An active fraction bias adds to any luminosity bias present in the sample, and, if present, would contribute to either an upward or downward trend for the lowest mass bins in Figure \ref{f:dmbhbins}, depending on the active fraction as a function of \mbh. The bottom panel of Figure \ref{f:lobin} shows an increasing trend of \mbh\ for the lower-limit of the $7.0 <$ log $\mbh < 7.5$ mass bin as a function of redshift. (The lowest mass bin is too limited in its redshift range to make this analysis.) As shown in the top panel of Figure \ref{f:lobin}, the downward trend in \dmbh\ for this mass bin is due to a greater trend of increasing \sigthree\ as a function of redshift. This may be evidence of a negative active fraction bias mitigated by a positive luminosity bias. 

% lobin

%% DISCUSSION

\section{Discussion}\label{s:discuss}

\subsection{Evolution in the \mbhsigstar\ relationship for SDSS quasars}

Is there evidence of genuine evolution in the \mbhsigstar\ relationship or just the appearance of it? Only the highest-mass ($\mbh > 10^{9.0}$ \msun) and lowest-mass ($\mbh < 10^{7.5}$ \msun) bins in Figure \ref{f:dmbhbins} show any change in \dmbh\ with redshift. Without knowing the magnitude of the intrinsic scatter or the shape of the scatter function for the quasar \mbhsigstar\ relationship, it is not possible to determine with accuracy to what degree all of these biases affect our sample. However, the observational biases and uncertainties described in \S \ref{s:biases} can account for most or all of the apparent evolution in these mass bins. The positive trend for the highest-mass bin is consistent with both a Lauer selection bias and a Shen-Kelly mass-uncertainty bias, while the negative trend for the lowest-mass bin is possibly due to a combination of a Schulze-Wisotzki active-fraction bias and a Malmquist-like luminosity bias in which lower-luminosity quasars (and therefore lower-mass black holes) are increasingly unlikely to be included in the sample with increasing redshift. The \mbhsigstar\ relationship for black holes in the middle mass range ($10^{7.5} < \mbh < 10^{9.0}$ \msun) shows minimal evolution with redshift. Inspection of the overall results suggests a limit of $\pm$0.2 dex on any evolution in the \mbhsigstar\ relationship for quasars from $z = 0.1$ to $z = 0.8$ compared with the relationship observed in the local universe. 

\subsection{Reconciling disparate results} 

The various studies of evolution in black hole -- galaxy relationships (see \S \ref{s:intro}) have painted a confused picture, with the conclusions spanning the range of possible results. Can these be reconciled into a single, coherent picture of black hole -- galaxy evolution? 

Lauer et al. (2007) cautioned that comparing AGN samples with the local sample of quiescent galaxies could produce a false signal of evolution because of the differing selection criteria used. Comparing one AGN sample with another AGN sample can also be problematic unless the selection criteria for both are the same. One way to mitigate this problem is to compare lower-$z$ and higher-$z$ samples that were selected using the same criteria, which we have done to some extent, with the added feature of the mass bins. As seen in Figure \ref{f:dmbhbins}, the magnitude and sign of the offset from the local relationship is a function of the black hole mass. This may provide some insight into the problem of the seemingly disparate results in studies of black hole - galaxy evolution. Many of the AGN studies of evolution focus on black hole -- galaxy relationships not only for relatively small redshift ranges, but for a relatively narrow range of black hole masses. It is possible that some of these studies are simply observing the offset of samples from the local relationship that arises from inherent scatter, as opposed to actual evolution. %While certainly not conclusive, Figure \ref{f:dmbhbins} presents an intriguing possibility for reconciling at least some of the seemingly disparate results. 

%% SUMMARY

\section{Summary}\label{s:summary}

We have investigated the \mbhsigstar\ relationship for SDSS DR7 quasars in the redshift range $0.1 < z < 1.2$ to assess whether any evolution has taken place. We used the ``photoionization method'' with the widths of the broad \hbeta\ or \mgii\ emission lines and the quasar continuum luminosity to calculate black hole masses, and the widths of the narrow \oiii\ or \oii\ lines as surrogates for the stellar velocity dispersion. We divided our lower-redshift HO3 sample into black hole mass bins incremented by 0.5 dex to assess the change in the \mbhsigstar\ relationship as a function of redshift for different black hole masses. All but one of the mass bins are offset from the local relationship, with the magnitude and direction of the offset determined by the black hole mass and the degree of scatter in the relationship. The offsets for the lowest-mass and highest-mass black holes appear to evolve with redshift; however, various observational biases can account for most or all of the apparent evolution. The offsets for the middle three mass bins remain approximately constant with redshift and suggest a limit of $\pm$0.2 dex on any evolution in the \mbhsigstar\ relationship up to $z = 0.8$. Finally, it may be possible to reconcile some of the seemingly disparate results from various studies of evolution in the black hole - galaxy relationships by accounting for offsets in the relationship for different masses due to scatter. 

\acknowledgments

We thank Mark Bottorff for helpful discussions and Alyx Stevens for assistance. G.A.S. gratefully acknowledges the hospitality of Lick Observatory and the support of the Jane and Roland Blumberg Centennial Professorship in Astronomy.

Funding for the Sloan Digital Sky Survey (SDSS) has been provided by the Alfred P. Sloan Foundation, the Participating Institutions, the National Aeronautics and Space Administration, the National Science Foundation, the U.S. Department of Energy, the Japanese Monbukagakusho, and the Max Planck Society. The SDSS Web site is http://www.sdss.org/. The SDSS is managed by the Astrophysical Research Consortium (ARC) for the Participating Institutions. The Participating Institutions are The University of Chicago, Fermilab, the Institute for Advanced Study, the Japan Participation Group, The Johns Hopkins University, the Korean Scientist Group, Los Alamos National Laboratory, the Max-Planck-Institute for Astronomy (MPIA), the Max-Planck-Institute for Astrophysics (MPA), New Mexico State University, University of Pittsburgh, University of Portsmouth, Princeton University, the United States Naval Observatory, and the University of Washington.

%% BIBLIOGRAPHY

\end{document}